%%%%%%%%%%%%%%%%%%%%%%%%%%%%%%%%%%%%%%%%%%%%%%%%%%%%%%%%%%%%%%%%%%%%%%%%%%%
%%%%%%%%%%%%%%%%%%%%%%%%%%%%%%%%%%%%%%%%%%%%%%%%%%%%%%%%%%%%%%%%%%%%%%%%%%%
%%%%%%%%%%%
%%%%%%%%%%%  Strong Coupling Expansions of SU(N) Seiberg-Witten Theory %%%%
%%%%%%%%%%%     
%%%%%%%%%%%                Eric D'Hoker and D.H. Phong                 %%%%
%%%%%%%%%%%
%%%%%%%%%%%%%%%%%%%%%%%%%%%%%%%%%%%%%%%%%%%%%%%%%%%%%%%%%%%%%%%%%%%%%%%%%%%
%%%%%%%%%%%%%%%%%%%%%%%%%%%%%%%%%%%%%%%%%%%%%%%%%%%%%%%%%%%%%%%%%%%%%%%%%%%
% 
%
\magnification=1200
\baselineskip=14pt
\overfullrule=0pt
\def\d{{\rm d}}
\def\12{{1 \over 2}}
\def\d{{\rm d}}

\def\l{\lambda}
\def\ln{{\rm log\,}}
\def\D{\partial}

\def\F{{\cal F}}

\def\L{\Lambda}

\def\+{{\rm Res}_{P_+}(z\d\l)}
\def\-{{\rm Res}_{P_-}(z\d\l)}
\bigskip
\rightline{IASSNS-HEP-97/2}
\rightline{UCLA/97/TEP/1}
\rightline{Columbia/Math/97/1}
\bigskip
\centerline{\bf STRONG COUPLING EXPANSIONS OF SU(N)}
\medskip
\centerline{{\bf SEIBERG-WITTEN THEORY}\footnote*{Research supported 
in part by
the National Science Foundation under grants PHY-95-31023 and DMS-95-05399,
and by the Harmon Duncombe Foundation.}}
\bigskip
\bigskip
\centerline{{\bf Eric D'Hoker}${}^1$ 									
            {\bf and D.H. Phong}${}^2$}
\bigskip
\centerline{${}^1$ Institute for Advanced Study}
\centerline{Princeton, NJ 08540, USA}
\centerline{and}
\centerline{Department of Physics}
\centerline{University of California, Los Angeles, CA 90024, USA}
\bigskip 
\centerline{${}^2$ Department of Mathematics}
\centerline{Columbia University, New York, NY 10027, USA}
\bigskip

\centerline{\bf Abstract}
\medskip
We set up a systematic expansion of the prepotential for ${\cal N}=2$ 
supersymmetric
Yang-Mills theories with SU(N) gauge group in the region of strong coupling
where a maximal number of mutually local fields become massless. 
In particular,
we derive the first non-trivial non-perturbative correction, 
which is the first 
term
in the strong coupling expansion providing a coupling between the different
U(1) factors, and which governs for example 
the strength of gaugino Yukawa couplings.

\vfill\break

\centerline{\bf I. INTRODUCTION}
\bigskip

The leading low energy behavior of ${\cal N}=2$ supersymmetric 
Yang-Mills theory 
is
completely determined by the effective prepotential $\F$, 
as a function of the
moduli of inequivalent vacuum states. The effective prepotential $\F$ is
obtained in terms of the periods of a meromorphic Abelian differential
$\d\lambda$, on a family of auxiliary Riemann surfaces parametrized by the 
moduli
of vacua. This construction was pioneered in [1] for gauge group SU(2), 
extended
in [2] to SU(N), to the inclusion of hypermultiplets in [3] 
and to other gauge
groups in [4]. The power of this result resides in the fact that it is 
valid for
all ranges of the coupling constant, including strong coupling.
Of special interest is the dynamics of the theory close to the
singularities in the moduli space of vacua; it is there that extra massless
degrees of freedom appear. 

At weak coupling, the Seiberg-Witten Ansatz reproduces the well-known
perturbative expression, together with multi-instanton corrections. 
For gauge
group SU(2), an expansion is readily derived from [1], and was found to agree
with direct calculations in field theory in [5] to the orders 
available (up to
field redefinitions). For gauge groups SU(N), SO(N) and Sp(N), (and any 
number of
hypermultiplets in the fundamental repesentation of the gauge group),
efficient methods for obtaining the instanton corrections to arbitrary
order were developed and 1- and 2-instanton contributions evaluated
explicitly in [6].
Geometrically, in the weak coupling region, the family of auxiliary Riemann
surfaces degenerates to a product of two spheres joined by N thin tubes, thus
producing an instanton expansion in terms of rational functions.

At strong coupling, singularities in the prepotential arise along
subvarieties in the moduli space of vacua where certain magnetic
magnetic monopoles or dyons become massless [1]. In view of the mass 
formula for
BPS states, such massless states will appear whenever one (or more) of the
periods of the differential $\d\lambda$ vanish. Geometrically, the strong
coupling singularities will occur at the subvarieties in the moduli
space of vacua where the auxiliary Riemann surfaces develop nodes. 

The charge assignments of the magnetic monopoles or dyons that become
massless at a given node, may be obtained directly from the
Picard-Lefschetz theorem, and their associated monodromies were found to
pass various physical consistency checks [1-4]. Their values may be
used to obtain the logarithmically singular contributions to the
prepotential, associated with the renormalization group behavior of
the infrared free light dyons at those points. For gauge groups of rank 
at least 2, Argyres and Douglas [7] found that special subvarieties occur
in the moduli space of vacua where mutually non-local dyons become 
massless simultaneously.

A more systematic analysis of the prepotential around strong coupling
singularities has been difficult to obtain in general, however. 
One method makes
use of the Picard-Fuchs equations for the periods of the meromorphic 1-form
$\d\lambda$. It concisely reproduces the results for gauge group SU(2), and
leads to explicit expressions for the quantum order parameters for SU(3) in
terms of Appell functions, from which a strong coupling expansion may be
obtained [8]. For gauge groups of rank 3, the associated 
Picard-Fuchs equations
were derived only recently, and their complexity and sheer 
size suggest that it
will be virtually impossible to generalize the results of [8] to gauge groups
of higher rank.

In the present paper, we shall set up methods by which systematic strong
coupling expansions of the prepotential at strong coupling 
singularities may be
obtained directly from the geometry of the Riemann surface and the meromorphic
1-form. 
The key ingredients are as follows : (1) the prepotential is easily
obtained from a renormalization group equation derived in [9-10], 
(2) the ''beta-function", i.e. the right hand side of the 
renormalization group,
is obtained by residue calculations on the small cycles of 
the Riemann surface.

We study theories with gauge group SU(N) and without hypermultiplets, but
the generalization to other gauge groups and the inclusion of 
hypermultiplets is
analogous. We shall concentrate on the region in the moduli space of 
vacua where
the maximal number of mutually local dyons become massless (which is N-1 for
gauge group SU(N)). These are precisly the strong coupling singularities for
which the auxiliary Riemann surface degenerates to two spheres connected by N
thin tubes, just as was the case for weak coupling, so that the 
prepotential may
be expanded in a series of rational functions. These maximal strong coupling
singularities are also the ones whose dynamics was studied in [11], in part in
the limit of large numbers of colors.
In particular,
we derive the first non-trivial non-perturbative correction, 
which is the first 
term
in the strong coupling expansion providing a coupling between the different
U(1) factors, and which governs for example 
the strength of gaugino Yukawa couplings.         

\bigskip
\bigskip

\centerline{\bf II. EXPANSION AROUND CURVES OF MAXIMAL SINGULARITY}

\bigskip

The renormalized order parameters $a_k$, their duals $a_{Dk}$ and the
prepotential are defined by
$$
2\pi i a_k=\oint_{A_k}\d\l,
\qquad
2 \pi i a_{Dk}= \oint_{B_k}\d\l 
\qquad
{\D\F(a_D)\over\D a_{Dk}}=a_{k}
\eqno (2.1)
$$ 
in terms of the meromorphic Abelian differential $d\l$, given by
$$
d\l = {x  A'(x) dx \over \sqrt {A(x)^2 - 4 \Lambda ^{2N}}}
\eqno (2.2)
$$
Here $A(x)$ is a polynomial of degree N in $x$, whose coefficients vary with
the moduli of vacua, and $\L $ is the renormalization scale. The associated
Riemann surface (or curve) is of genus $N-1$ and is given by 
$y^2 = A(x)^2 - 4 \L
^{2N}$. Without loss of generality (but possible rescaling of $\L$), the
polynomial $A$ may be normalized  so that $A(x) = x^N +  u x^{N-2} + {\cal O}
(x^{N-3})$. 

We shall make use of the renormalization group equation for the prepotential
found in [8,9] and expressed here in terms of the dual order parameters :
$$
\sum _{k=1} ^N a_{Dk} {\D \F \over \D a_{Dk} } - 2 \F 
  = {2N \over 2 \pi i}\ u  (a_D)
\eqno (2.3)
$$
where $u$ is the expansion coefficient of $A(x)$ as defined above.
This equation is extremely useful, because it now suffices to obtain the
coefficient $u$ as a function of the periods $a_{Dk}$. When the $B_k$ cycles
degenerate, such a result is easily gotten from residue calculations, as
was shown for the weak coupling regime in [10]. On the other hand, obtaining 
$\F$ directly from the the dual $A_k$ cycles, of (2.1), would involve
complicated integrals over long cycles. Notice that in the above equation 
(2.3),
the r\^oles of the renormalized order parameters $a_k$ and of their duals
$a_{Dk}$ may be interchanged.

\bigskip

\noindent
{\bf (a) The maximal singularity curves}

The strong coupling singularities at which a maximal number of mutually local
magnetic monopoles or dyons become massless are characterized by curves
$y^2 = A_0(x)^2 - 4 \L ^{2N}$ with a maximal number of distinct 
double zeros of
$y^2$, and thus no zeros of order higher than 2. (Zeros of order 3 and above
produce Argyres-Douglas singularities, and we shall not sudy those here.) As
was shown in [11], these curves are given by the following (Chebychev)
polynomials
$$
A_0 (x) = 2 \L ^N C_0 ({x \over 2 \L})
\qquad \qquad 
C_0 (z) = \cos \bigl ( N \arccos (z) \bigr )
\eqno (2.4)
$$
Since only $\L^{2N}$ entered the original problem, and $A_0(x)$ is a function
involving only $\L^2$, there are in general $N$ different solutions $A_0(x)$,
associated with the N different roots of unity which can multiply $\L$. All
these solutions are physically equivalent and we shall restrict to the one for
which $\L$ is real. 

The neighborhood of the maximal singularity curve may be parametrized by the
$N-1$ parameter family of polynomials $P$ of degree $N-2$, whose coefficients
are assumed to be small in units of $\L$. Without loss of generality, we may
render its coefficients dimensionless by scaling out suitable factors of $\L$,
similar to the form of (2.4). We then obtain the following parametrization 
of the
neighborhood of the maximal singularity
$$
A(x) = A_0(x) + 2 \L ^N P({x \over 2\L})
\eqno (2.5)
$$
Rescaling $x$ by a factor of $2 \L$ so that it too becomes
dimensionless, we have
$$
d\l
 =  2 \L { x (C_0 '(x) + P'(x) ) dx \over \sqrt {(C_0(x) + P(x))^2 - 1}}
\eqno (2.6)
$$
Henceforth, we shall measure the renormalized order parameters $a_k$ and their
duals $a_{D,k}$ in units of $2\L$ and thus set $\L = 1/2$.

The double zeros of $C_0 (x)$ are given by
$$
c_k = \cos \bigl ( { k \pi \over N} \bigr )
\qquad \qquad
k=1, \cdots , N-1
\eqno (2.7)
$$
The coefficients of the polynomial $P(x)$ parametrize the way in which the
$N-1$ double zeros are being approached. Generically, the effect of adding 
$P(x)$
in (2.5) is to lift each double zero to a small branch cut in the double
sheeted plane representation of the curve. We define the degenerating cycle
$B_k$, $k=1, \cdots , N-1$, as the cycle that surrounds the branch cut created
from the double zero $c_k$ once in the counterclockwise direction. The cycles
$A_k$ can then be chosen to be any set of conjugate cyles, but we shall not
need them here.

\bigskip

\noindent
{\bf (b) Expansion around degenerating cycles}

We assume that the coefficients of $P(x)$ are small compared to 1, 
and that the
curves $B_k$ are at a distance from $c_k$ of order a fraction of 1, (and thus
much larger than the coefficients of $P(x)$). Then, we may expand the period
integrals for the $a_{Dk}$ periods associated with the degenerating 
cycles $B_k$ 
in
a convergent power series of $P(x)$. The radius of convergence for this
expansion can be shown to be of order a fraction of 1 in these units.

To perform this expansion in an efficient way, it is most convenient to
introduce an auxiliary parameter $\mu$ which interpolates between $C_0$ and
$C_0 +P$ by $C_\mu (x) = C_0(x) + \mu P(x)$. The period integrals now become
$$
a_{Dk} = - \int _0 ^1 d\mu \oint _{B_k} {dx \over 2 \pi i} 
      {P(x) \over \sqrt{ C_\mu (x)^2 - 1}} 
=\sum _ {m=0 } ^\infty a_{Dk} ^{(m)} 
\eqno (2.8)
$$
with the expansion terms $a_{Dk} ^{(m)}$ given by
$$
a_{Dk} ^{(m)} = 
(-1)^{m+1} {\Gamma (m+ \12) \over \Gamma (\12) m!}
\int _0 ^1 d\mu  \mu ^m \oint _{B_k} {dx \over 2 \pi i} 
{P(x)^{m+1} (\mu P(x) + 2 C_0(x) )^m \over (C_0 (x) ^2 -1) ^{m+\12}}
\eqno (2.9)
$$
We fix the square root sign by  $ (C_0(x) ^2 -1) ^{\12}
= - i \sin (N \arccos x)$.  

{} For low rank $N-1$ of the gauge group, the polynomial $P(x)$ will be
of low degree, and one may use the coefficients of the polynomial as suitable
classical parameters of moduli space of vacua. As soon as $N$ becomes larger,
this parametrization is not convenient however. To identify a good
parametrization, we analyze the first term ($m=0$) in the above expansion.
We introduce the very convenient notation
$$
P_k = P(c_k); 
\qquad \qquad
s_k = \sin ( {k \pi \over N}), 
\qquad  
c_k = \cos ( {k \pi \over N}),
\qquad
k=1,\cdots , N-1
\eqno (2.10)
$$
The first expansion term is given by 
$$
a_{Dk} ^{(0)} =  i (-)^k {s_k \over N} \ P_k
\eqno (2.11)
$$
which indicates that convenient local coordinates of the moduli space of
vacua around the maximal singularity are the values $P_k$ of $P(x)$ at 
the double
points. 
We recover $P(x)$ with the help of the Lagrange interpolating formula giving 
the polynomial of degree $N-2$ in terms of its $N-1$ values $P_k$ at 
$N-1$ fixed
points $c_k$
$$
P(x) = \sum _{k=1} ^{N-1}  { C_0 '(x) P_k \over (x-c_k) C_0 ''(c_k)} 
\eqno (2.12)
$$
To obtain useful formulas for the expansion coefficients, we take advantage of
the fact that the factor of $C_0'(x)$ almost eliminates the denominator in
the integrand of (2.9). More precisely, we have 
$$
{C_0'(x) \over \sqrt {C_0(x)^2 -1} } = {iN \over \sqrt {1-x^2}}
\eqno (2.13)
$$
We now introduce the functions $\Pi _k(x)$ and $G_k(x)$, which are
analytic at $x=c_k$; we also define the functions $\Pi(x)$ and
$G(x)$, which have simple poles at all $c_k$, and which we shall make
use of later on.
$$
\eqalign{
\Pi _k (x) &= (x-c_k) \Pi (x) 
\qquad \Pi(x) = N^2 {P(x) \over C_0 '(x)} 
           = - \sum _{l=1} ^{N-1}  (-) ^l {P_l s_l^2 \over x-c_l} 
\cr
G _k (x) & = (x-c_k)  G (x)
\qquad  G(x) = N^2 { C_0(x) \over C_0 '(x)}
           =  xN - \sum _{l=1} ^{N-1} { s_l ^2 \over x-c_l}
\cr}
\eqno (2.14)
$$ 
In terms of these functions, the expansion coefficients $a_{Dk}^{(m)}$ 
of (2.9)
are easily evaluated, and we obtain
$$
\eqalign{
a_{Dk} ^{(m)} 
=  &
- i{\Gamma (m+ \12)/\Gamma (\12) \over  m!  N^{2m+1} }
\int _0 ^1 d\mu  \mu ^m  \oint _{B_k} {dx \over 2 \pi i}
{(1-x^2)^{-m- \12}  \over (x-c_k)^{2m+1}}
\Pi _k (x)^{m+1} (\mu \Pi _k (x) + 2 G_k (x) )^m  
\cr
& \cr
= &
- i{\Gamma (m+ \12)/\Gamma (\12) \over  m! (2m)!  N^{2m+1} }
\int _0 ^1 d\mu  \mu ^m  {\D^{2m} \over \D x^{2m}} \bigg | _{x=c_k}
\! \! \! (1-x^2)^{-m- \12}  
\Pi _k (x)^{m+1} (\mu \Pi _k (x) + 2 G_k (x) )^m  
\cr}
\eqno (2.15)
$$
Here, the $x$-integral was evaluated by residue methods. Formula (2.15)
provides an expansion to all orders in $P_k$ of the dual quantum order 
parameters
$a_{Dk}$. To evaluate the expansion coefficients, it will prove convenient to 
have the following derivative terms
$$
\eqalign{
G_k (c_k)     = & - s_k ^2  \cr
G_k ' (c_k)   = & {3\over 2} c_k \cr
G_k ''(c_k)   = & ({2 \over 3} N^2 s_k ^2  
+ {4 \over 3} s_k ^2 + \12 c_k ^2)
                {1 \over s_k ^2} \cr
G_k ''' (c_k) = & ( - N^2 s_k ^2 + {3\over 2} c_k ^2 + {7 \over 4} s_k ^2)
              {c_k \over s_k ^4} \cr}
\eqno (2.16)
$$
as well as the fact that $\Pi _k (c_k) = - (-)^k s_k ^2 P_k$.

\bigskip

\noindent
{\bf (c) Contributions to second order}

We shall now use the above expression to evaluate contributions to second order
in $P_k$ first, and find that
$$
\eqalign{
a_{Dk} ^{(1)}  \big |_{{\rm quadratic \ in} \ P_k } 
 =
-{ i \over 2 s_k N^3 } \bigl [ &
(-{5 \over 4} c_k ^2 + {2 \over 3} s_k ^2 - {3 \over 2} + {1 \over 3} N^2 s_k
^2) P_k ^2  -    \Pi _k '(c_k) ^2 \cr
&
 + (-)^k s_k ^2  P_k \Pi ''_k (c_k) 
  + 3(-)^k c_k   P_k \Pi _k '(c_k)  \bigr ]
\cr}
\eqno (2.17)
$$
The derivatives are easily evaluated and given as follows
$$
\eqalign{
\Pi _k ' (c_k) =& - \sum _{l\not=k} ^{N-1} (-)^l 
{ s_l ^2 \over c_k - c_l} P_l
\cr
\Pi _k '' (c_k) =& 2 \sum _{l\not=k} ^{N-1} (-)^l 
{ s_l ^2 \over (c_k - c_l)^2}
P_l
\cr}
\eqno (2.18)
$$
Higher order terms may be calculated analogously, but we shall not need them
here. Instead, we shall use some shortcuts and compute directly the beta
function.

\bigskip
\bigskip

\centerline{{\bf III. THE BETA FUNCTION AND THE PREPOTENTIAL}}

\bigskip

The renormalization group beta function $u$, discussed in (2.3) is 
obtained in terms of the leading coefficient multiplying $x^{N-2}$ of $P(x)$.
Remarkably, this quantity also assumes a simple expression in terms of the
coordinates $P_k$,  and we find
$$
u =   - 4 \L ^2 \tilde u ; 
\qquad \tilde u =  {1 \over N} \sum _{k=1} ^{N-1} (-)^k s_k ^2 P_k
\eqno (3.1)
$$
Here, we have dropped a constant contribution ($-N \L^2$) to $u$, since
it would yield just a constant shift of the prepotential $\F$ which is 
physically
immaterial.

Now, instead of calculating out the expansions of $a_{Dk}$ in terms of the
classical variables $P_k$, then inverting to obtain an expansion of $P_k$
in terms of $a_{Dk}$, and finally inserting the expansions of $P_k$ into  
(2.19)
to obtain the beta function $u$, it is much more advantageous to  exhibit the
summation that occurs in $\tilde u$ before effecting the substitution. In
particular, this procedure will avoid having to perform major resummations
which would have to be carried out by contour integrations anyway. 
Thus, we shall
instead consider the following sum which directly exhibits
$\tilde u$
$$
\tilde u = -i \sum _{k=1} ^{N-1} s_k a_{Dk}   +  \sum _{m=1} ^\infty
\alpha _D ^{(m)}
\qquad \quad
\alpha _D ^{(m)} = i \sum _{k=1} ^{N-1} s_k a_{Dk} ^{(m)}
\eqno (3.2)
$$ 
The resummed expansion parameters $\alpha _D ^{(m)}$ may be expressed 
in terms of
the functions $\Pi(x) $ and $G(x)$, defined in (2.14), and are given by
$$ 
\alpha _D ^{(m)} = \sum _{k=1} ^{N-1} 
 {\Gamma (m+ \12) \over \Gamma (\12) m!  N^{2m+1} }
\int _0 ^1 d\mu  \mu ^m  \oint _{B_k} {dx \over 2 \pi i} \ 
{ s_k \over (1-x^2)^{m+ \12} } \ 
\Pi  (x)^{m+1} (\mu \Pi  (x) + 2 G (x) )^m  
\eqno (3.3)
$$
What prevents us from carrying out the sum over $k$ by connecting the 
various contours $B_k$ in the above formula is twofold. 
First, the square root
would force us to perform a line integration around the lines from $\pm1$ to
$\infty$. This would not represent an advantage over a direct evaluation
with residue methods. Second, the factor $s_k$ depends upon $k$, so we cannot
just add up the various contours $B_k$.

\bigskip

\noindent
{\bf (a) Combining contours}

We can solve both problems by replacing $s_k $ inside the 
above integrand by a
$k$-independent function that takes the value $s_k $ at 
$x=c_k$ for all
$k=1,\cdots,N-1$, but whose derivatives of orders 
$1,\cdots, 2m$ vanish at this
point. Replacing $s_k$ by such a function inside (3.3) 
for each value of $m$
will not alter the value of the integral. It is very easy 
to obtain such a
function by expanding the value at $c_k$ in a Taylor series 
around $x$ instead.
For example, defining $s(x) = \sqrt{1-x^2}$,  we have
$$
s_k = s(c_k) = s(x) + (c_k -x) s'(x) + \12 (c_k -x)^2 s''(x) + {\cal
O}((c_k-x)^2)
\eqno (3.4)
$$ 
To order $m=1$, the expressions we need are given by
$$
{s_k \over (1 - x^2 ) ^{3/2}} = {1 \over 1-x^2} + (x-c_k) 
{x \over (1 - x^2 )^2}
- \12 (x- c_k )^2 {1 \over (1-x^2)^3} + {\cal O}((x-c_k )^3)
\eqno (3.5)
$$
while to order $m=2$, we have 
$$
\eqalign{
{s_k \over (1 - x^2 ) ^{5/2}} G(x)^2 = & {G(x)^2 \over (1-x^2)^2} - {x \over
(1-x^2)^2} G(x) - \12 {1 \over 1-x^2} -{ c_k \over 6 s_k ^4} 
(1+ 2N ^2) (x-c_k)
\cr
&
-{1 \over 24 s_k ^6} (3 + 11 c_k ^2 + 28 N^2 c_k ^2) (x- c_k )^2 + {\cal
O}((x-c_k)^3) \cr}
\eqno (3.6)
$$
Inside the integrals, the ${\cal O}()$ terms vanish identically and may be
dropped. What we have gained is that the first term in (3.5) 
and the first three
terms in (3.6) are $k$-independent and various contours in the sum 
(3.3) may now
be collected together. As can be established by inspecting (3.3), (3.5) and
(3.6), the only poles other than those occuring at 
$c_k, \ k=1,\cdots, N-1$ are
the poles at $x=\pm 1$. Thus, we have for those parts of the integral
$$
\sum _{k=1} ^{N-1} \oint _{B_k} = - \oint _{\pm 1}
\eqno (3.7)
$$
The remaining terms are handled by standard residue calculations.

\bigskip

\noindent
{\bf (b) Contributions for $m=1$}

Substituting (3.5) into (3.3) for $m=1$, we find, using also (3.7)
$$
\eqalign{
\alpha _D ^{(1)} = 
&  -{1 \over 2N^3} \oint _{\pm 1} {dx \over 2 \pi i}
   {1 \over 1-x^2} \bigl ( {1 \over 3} \Pi (x) ^3 + \Pi (x)^2 G(x) \bigr )
\cr
& + {1 \over 2N^3} \sum _{k=1} ^{N-1} \oint _{B_k} {dx \over 2 \pi i}
     \{x(1-x^2) -{1 \over 2} (x-c_k)\}
    {\ {1 \over 3} \Pi _k (x)^3 + \Pi _k (x) ^2 G_k (x) \ \over (x-c_k)^2
     (1-x^2)^3}
\cr}
\eqno (3.8)
$$
It is straightforward to carry out the residue calculations at $\pm 1$ in the
first integrals, and at $c_k, \ k=1,\cdots, N-1$ in the second integrals.
Using also the fact that $G(\pm 1) =\pm 1$, we find
$$
\eqalign{
\alpha _D ^{(1)} = 
&
    +{1 \over 4 N^3} \bigl \{ \Pi(1)^2 + \Pi (-1)^2 + {1 \over 3} \Pi (1)^3
    - {1 \over 3} \Pi(-1)^3 \bigr \}
\cr
&
    -{ 1 \over 4 N^3} \sum _{k=1} ^{N-1} \biggl \{ (1 + 3 c_k ^2) P_k ^2
    -4  (-) ^k c_k P_k \Pi _k ' (c_k)  + (-)^k ({1 \over 3} 
    + 2 c_k ^2) P_k ^3
    -2 c_k P_k ^2 \Pi _k '(c_k) \biggr \}
\cr}
\eqno (3.9)
$$
Notice that the combinations $\Pi (\pm 1)$ are simply given by
$$
\Pi(\pm 1) = - \sum _{k=1} ^{N-1} (-)^k (c_k \pm 1) P_k
\eqno (3.10)
$$

Making use of expression (2.18) for $\Pi_k '(c_k)$, and some straightforward
trigonometric identities, we find
$$
\eqalign{
\alpha _D ^{(1)} =
& {1 \over 4 N^3} \sum _{k=1} ^{N-1} s_k ^2 P_k^2
    \ + {1 \over 12 N^3}  (   \Pi (1)^3 -  \Pi(-1)^3  )
\cr
&    -{ 1 \over 4 N^3} \sum _{k=1} ^{N-1} \bigl \{ 
   (-)^k ({1 \over 3} + 2 c_k ^2) P_k ^3
    -2 c_k P_k ^2 \Pi _k '(c_k) \bigr \}
\cr}
\eqno (3.11)
$$
Notice that the quadratic part is diagonal in $P_k$.

\bigskip

\noindent
{\bf (c) Contributions for $m=2$}

In a manner analogous to the computation of the $m=1$ contributions, we get
those for $m=2$ by substituting (3.6) into (3.3) for $m=2$. Using the fact
that $G'(\pm 1) = {1 \over 3} (2N^2 +1)$, we find 
$$
\eqalign{
\alpha _D ^{(2)} = 
&  - {1 \over 24 N^5} (2N^2 +1) ( \Pi (1)^3 - \Pi (-1)^3) 
\cr
&
-{1 \over N^5} \sum _{k=1} ^{N-1}
\biggl \{
c_k ({1 \over 4} + \12 N^2) P_k ^2 \Pi _k '(c_k)
-(-)^k ({1 \over 16} + {11 \over 48} c_k ^2 +{7 \over 12} N^2 c_k ^2) P_k ^3
\biggr \} 
\cr}
\eqno (3.12)
$$

\bigskip

\noindent
{\bf (d) Expressing $P_k$ and the beta function in terms of $a_{Dk}$}

To obtain the beta function to cubic order in $a_{Dk}$, it suffices 
to substitute 
into (3.11) and (3.12) the expansion of $P_k$ in terms of $a_{Dk}$ up to 
quadratic order only. This inversion is easily obtained from (2.11) 
and (2.17),
and we find
$$
\eqalign{
P_k = - (-)^k {iN \over s_k} a_{Dk} - { (-)^k \over 2 s_k ^4} \bigl [
&
(-{5 \over 4} c_k ^2 + {2 \over 3} s_k ^2 - {3 \over 2} 
+ {1 \over 3} N^2 s_k
^2) a_{Dk} ^2 + {s_k ^2 \over N^2} \Pi _k '(c_k) ^2
\cr
& 
+ i {s_k ^3\over N} a_{Dk} \Pi _k ''(c_k) 
+ {3i \over N} s_k c_k a_{Dk} \Pi _k ' 
(c_k)
\bigr ] + {\cal O}(a_{Dk} ^3)
\cr}
\eqno(3.13)
$$
Putting all together, we find for the beta function the following expression
$$
\eqalign{
\tilde u = &
 -i \sum _{k=1} ^{N-1} s_k a_{Dk} 
 - { 1\over 4N} \sum _{k=1} ^{N-1} a_{Dk}^2 
 - {1 \over 24 N^5}  (\Pi (1)^3 - \Pi (-1)^3)
\cr
&
-{i \over 4 N^2} \sum _{k=1} ^{N-1} { 1 \over s_k ^3} 
   \biggl [ 
   ( {7 \over 12} + c_k ^2 ) a_{Dk} ^3 
   - {s _k ^2 \over N^2} a_{Dk}  \Pi _k '(c_k) ^2
   - i {s_k ^3 \over N} a_{Dk} ^2 \Pi _k '' (c_k)
   - {2i \over N}  c_k s_k a_{Dk} ^2 \Pi _k ' (c_k)
   \biggr ]
\cr}
\eqno (3.14)
$$
Here, the derivatives of $\Pi _k$ are evaluated as follows
$$
\eqalign{
\Pi _k ' (c_k) = 
&
iN \sum _{l\not=k} { s_l a_{Dl} \over c_k - c_l}
\cr
\Pi _k '' (c_k) = 
&
-2iN \sum _{l\not=k} { s_l a_{Dl} \over (c_k - c_l)^2}
\cr}
\eqno (3.15)
$$
It remains to rearrange the sum in the last line of (3.14), and 
to combine with the $\Pi(+1) ^3 - \Pi (-1)^3$ term of the first
line. The result is as follows
$$
\tilde u = 
 -i \sum _{k=1} ^{N-1} s_k a_{Dk} 
 - { 1\over 4N} \sum _{k=1} ^{N-1} a_{Dk}^2 
 -{i \over 16 N^2}
 \biggl [ \sum _{k=1} ^{N-1} {a_{Dk} ^3 \over s_k ^3}
 -4 \sum _{k\not=l} ^{N-1} { a_{Dk} ^2  a_{Dl} s_l \over (c_k - c_l)^2}
 \biggr ]
\eqno (3.16)
$$
Restoring the original $\Lambda$-dependence, we now easily find
the renormalization group beta function
$$ 
 u = 
 2i\Lambda  \sum _{k=1} ^{N-1} s_k a_{Dk} 
 + { 1\over 4N} \sum _{k=1} ^{N-1} a_{Dk}^2 
 +{i \over 32 N^2 \Lambda } 
 \biggl [
 \sum _{k=1} ^{N-1} {a_{Dk} ^3 \over s_k ^3}
 -4 \sum _{k\not=l} ^{N-1} { a_{Dk} ^2  a_{Dl} s_l \over (c_k - c_l)^2}
 \biggr ]
\eqno (3.17)
$$

\bigskip

\noindent
{\bf (e) The prepotential}

Starting from the result for the beta function in (3.17), 
and making use of (2.3), we obtain our final expression for 
the prepotential at strong coupling
$$ 
\eqalign{
 {\cal F} (a_D) = &
 -{2 N \Lambda \over \pi}   \sum _{k=1} ^{N-1} s_k a_{Dk} 
 - { i\over 4\pi } \sum _{k=1} ^{N-1} a_{Dk}^2 \ln {a_{Dk} \over \Lambda _k} 
 \cr
 & \qquad \qquad
 +{1 \over 32 \pi N \Lambda } 
 \biggl [
 \sum _{k=1} ^{N-1} {a_{Dk} ^3 \over s_k ^3}
 -4 \sum _{k\not=l} ^{N-1} { a_{Dk} ^2 a_{Dl} s_l\over (c_k - c_l)^2}
 \biggr ]
 \cr}
\eqno (3.18)
$$
Here, the values of the normalizations $\Lambda _k$
are not available directly from the integration of the 
renormalization group equation in (2.3). 
(In principle, the renormalization group equation will determine
the prepotential ${\cal F}$ through (2.3) only up to a homogeneous
function of degree 2 in $a_{Dk}$. Here, however, we are performing 
an expansion in powers of $a_{Dk} / \Lambda$, and we know what that
expansion looks like from the evaluation of the $A_k$ periods.
In particular, only a single logarithmic term will appear.
Thus, knowing the form of the expansions of both $u$ and ${\cal F}$,
the only remaining indeterminacy of the renormalization group 
equation are terms quadratic in $a_{Dk}$.) They can be obtained
from a direct calculation to that order of the $A_k$ periods, 
a calculation that has already been carried out by 
Douglas and Shenker [11]. We shall 
quote their results here :
$$
\ln {\Lambda _k \over \Lambda} = {3 \over 2} + \ln s_k
\eqno (3.19)
$$

\medskip

\noindent
{\bf (f) Comparison with known results}

{} For SU(2), N=2, we have $s_1=1$ and thus
$$
{\cal F} = - {4 \Lambda \over \pi} a _D 
           - { i \over 4 \pi} a_D ^2 \ln {a_D \over \Lambda}
           + {1 \over 64 \pi \Lambda} a _D ^3
\eqno (3.20)
$$
which agrees with [1,8] after suitably adjusting for the slightly different
definitions of $\Lambda$. For SU(3), N=3, we have $s_1=s_2=\sqrt{3}/2$
and $c_1=-c_2 = 1/2$ and thus
$$
\eqalign{
{\cal F} = & -{3 \sqrt{3} \Lambda \over \pi} (a_{D1} + a_{D2})
           -{i \over 4 \pi} (a_{D1} ^2 \ln {a_{D1} \over \Lambda _1}
                            +a_{D2} ^2 \ln {a_{D2} \over \Lambda _2})
                            \cr
                            &
           +{1 \over 144 \sqrt {3} \pi \Lambda} (a_{D1} + a_{D2}) 
                                           (4a_{D1} ^2 -13 a_{D1} a_{D2} + 4 
a_{D2} ^2)
                                           \cr}
\eqno (3.21)
$$
Again, taking into account the difference in definition of $\Lambda$
with [8], we find agreement. Finally, for all SU(N), the logarithmic
terms were computed in [11], and the normalization of the logs agrees.

\medskip

\noindent
{\bf (g) Physical interpretation}

To understand more clearly the physical origin of the non-perturbative
correction terms that we have calculated, it is helpful to derive, 
within this 
approximation,
the (dual gauge) coupling constant $\tau ^D _{mn} = \partial ^2 {\cal F} /
 \partial a_{Dm} \partial a_{Dn}$. We find
$$
\eqalign{
\tau ^D _{mm} = & - {i \over 2 \pi} \ln {a_{Dm} \over \Lambda _m} 
- {3i \over 4 
\pi}
             + {3 \over 16 \pi N \Lambda } { a_{Dm} \over s_m ^3}
             - {1 \over 8 \pi N \Lambda} \sum _{l\not=m} {2 a_{Dl} s_l 
             \over (c_m - c_l )^2}
             \cr
\tau _{mn} = & - {1 \over 4 \pi N \Lambda } { a_{Dm} s_n + a_{Dn} s_m \over
                 (c_m-c_n)^2}
\cr}
\eqno (3.22)
$$                              
Identifying $\tau ^D _{mm} $ in terms of the dual coupling constant 
$g_{Dm}$ and
dual instanton angle $\theta _{Dm}$,
as usual, we find that the cubic correction terms in the prepotential
have a dependence on $g_{Dm}$ and $\theta _{Dm}$ through 
$$
e^{2 \pi i \tau ^D _{mm}} = e^{- {8 \pi ^2 \over g_{Dm}^2} + i \theta _{Dm}}
\eqno (3.23)
$$ 
suggesting
that the effects are due to (magnetic sector) instantons. It would be very 
illuminating to see how the same effects can be identified from the
physics of light monopoles near the strong coupling points.

\bigskip
\bigskip

\centerline{\bf REFERENCES}
\bigskip

\item{[1]} N. Seiberg and E. Witten, Nucl. Phys. B 426 (1994) 19, 
hep-th/9407087;
\item{[2]} A. Klemm, W. Lerche, S. Yankielowicz, and S. Theisen,
Phys. Lett. B 344 (1995) 169;\hfil\break
P.C. Argyres and A. Faraggi, Phys. Rev. Lett. 73 (1995) 3931;
\item{[3]} N. Seiberg and E. Witten, Nucl. Phys. B 431 (1994) 484,
hep-th/9408099;\hfil\break
A. Hanany and Y. Oz, Nucl. Phys. {\bf B452} (1995) 73,
hep-th/9505075;\hfil\break
P.C. Argyres and A. Shapere, Nucl. Phys. B 461 (1996) 437,
hep-th/9609175;\hfil\break
A. Hanany, Nucl. Phys. B 466 (1996) 85, hep-th/9509176.
\item{[4]} P.C. Argyres, R. Plesser, and A. Shapere, Phys. Rev. Lett. 75
(1995) 1699, hep-th/9505100;\hfil\break
J. Minahan and D. Nemeshansky, Nucl. Phys. B 464 (1996)
3, hep-th/9507032;\hfil\break
U.H. Danielsson and B. Sundborg, Phys. Lett. B 358 (1995) 273;
hep-th/9504102;\hfil\break
A. Brandhuber and K. Landsteiner, Phys. Lett. B 358 (1995) 73, 
hep-th/9507008;
\hfil\break
M. Alishahiha, F. Ardalan, and F. Mansouri, Phys. Lett. B 381 (1996)
446, hep-th/9512005.
\item{[5]} N. Dorey, V. Khoze, and M. Mattis, Phys. Rev. D 54 (1996)
7832, hep-th/9606199; 
Phys. Lett. B 388 (1996) 324, hep-th/9607066; hep-th/9607202;\hfil\break
Y. Ohta, hep-th/9604051, hep-th/9604059;\hfil\break
D. Finnell and P. Pouliot, Nucl. Phys. B 453 (1995) 225;
\hfil\break
A. Yung, hep-th/9605096; \hfil\break
F. Fucito and G. Travaglini, hep-th/9605215\hfil\break
H. Aoyama, T. Harano, M. Sato, and S. Wada, hep-th/9607076;\hfil\break  
T. Harano and M. Sato, hep-th/9608060
\item{[6]} E. D'Hoker, I.M. Krichever, and D.H. Phong, hep-th/9609041;
hep-th/9609145, to appear in Nucl. Phys. B;
\item{[7]} P. Argyres and M. Douglas, Nucl. Phys. B 448 (1995) 93,
hep-th/9505062.
\item{[8]} A. Klemm, W. Lerche, and S. Theisen, Int. J. Mod. Phys. A 11
(1996) 1929, hep-th/9505150;\hfil\break
K. Ito and S.K. Yang, Phys. Rev. D 53 (1996) 2213,
hep-th/9603073;\hfil\break
J.M. Isidro, A. Mukherjee, J.P. Nunes, and H.J. Schnitzer,
hep-th/9609116.
\item{[9]} M. Matone, Phys. Lett. B 357 (1995) 342;\hfil\break
G. Bonelli and M. Matone, Phys. Rev. Lett. 76 (1996) 4107;\hfil\break
G. Bonelli and M. Matone, hep-th/9605090; \hfil\break
T. Eguchi and S.K. Yang, Mod. Phys. Lett. A 11 (1996) 131;
\hfil\break
P. Howe and P. West, hep-th/9607239 \hfil\break
J. Sonnenschein, S. Theisen, and S. Yankielowicz, Phys.
Lett. B 367 (1996) 145, hep-th/9510129.
\item{[10]} E. D'Hoker, I.M. Krichever and D.H. Phong, hep-th/9610156.
\item{[11]} M. Douglas and S. Shenker, Nucl. Phys. B 448 (1995) 
271, hep-th/9503163.

\end